\def\year{2022}\relax
\documentclass[letterpaper]{article} % DO NOT CHANGE THIS
\usepackage{aaai22}  % DO NOT CHANGE THIS
\usepackage{graphicx} % Required for inserting images
\usepackage[utf8]{inputenc}
\usepackage{pgf, tikz}
\usepackage{pgfplots}
\usepackage{adjustbox}
\usetikzlibrary{arrows, automata}
\usetikzlibrary{graphs, graphs.standard}
\usepackage{atbegshi}
\usepackage{amsmath}
\usepackage{ stmaryrd } 
\usepackage{ dsfont }
\usepackage[numbers]{natbib}
\usepackage{booktabs}
\usepackage{longtable}
\usepackage[hyphens]{url}  % DO NOT CHANGE THIS
\usepackage{graphicx} % DO NOT CHANGE THIS
\usepackage{footnote}
\usepackage{caption}
\usepackage{algorithm}
\usepackage{epstopdf}% To incorporate .eps illustrations using PDFLaTeX, etc.
\usepackage{algpseudocode}

\usepackage{amsthm}
\newtheorem{definition}{Definition}[section]

\usepackage{hyperref}
\hypersetup{
    colorlinks=true, %set true if you want colored links
    linktoc=all,     %set to all if you want both sections and subsections linked
    linkcolor=red,  %choose some color if you want links to stand out
}

\usepackage{newfloat}
\usepackage{listings}
\floatstyle{ruled}
\newfloat{listing}{tb}{lst}{}
\floatname{listing}{Listing}
\title{A Chip-Firing Game for Biocrust Reverse Succession}
\author {
    Shloka V. Janapaty \footnote{ORCID: 0000-0001-9697-0471}
}
\affiliations{
    Department of Applied Physics and Applied Mathematics\\
    Columbia University\\
    New York, NY, USA \\
}

\usepackage{bibentry}
\tikzset{
  common/.style={draw,name=#1,node contents={},inner sep=0,minimum size=3},
  disc/.style={circle,common=#1},
  square/.style={rectangle,common={#1}},
}

\providecommand{\keywords}[1]{\textbf{Keywords:} #1}
\begin{document}

\maketitle

\begin{abstract}
Experimental work suggests that biological soil crusts, dominant primary producers in drylands and tundra, are particularly vulnerable to disturbances that cause reverse ecological succession. To model successional transitions in biocrust communities, we propose a resource-firing game that captures succession dynamics without specifying detailed function forms. The game is evaluated in idealized terrestrial ecosystems, where disturbances are modeled as a reduction in available resources that triggers inter-species competition. The resource-firing game is executed on a finite graph with nodes representing species in the community and a sink node that becomes active when every species is depleted of resources. First, we discuss the theoretical basis of the resource-firing game, evaluate it in the light of existing literature, and consider the characteristics of a biocrust community that has evolved to equilibrium. We then examine the dependence of resource-firing and game stability on species richness, showing that high species richness increases the probability of very short and long avalanches, but not those of intermediate length. Indeed, this result suggests that the community’s response to disturbance is both directional and episodic, proceeding towards reverse succession in bursts of variable length. Finally, we incorporate the spatial structure of the biocrust community into a Cayley Tree and derive a formula for the probability that a disturbance, modeled as a random attack, initiates a large species-death event.
\end{abstract}


\section{Introduction}
Global change threatens to destabilize terrestrial ecosystems by triggering ecological state transitions, significant shifts in community structure and function. These shifts may occur in response to diminished resource availability or climate-induced disturbances, including increased frequency and intensity of extreme events, altered precipitation patterns, and rising temperatures. In some cases, changes in disturbance regimes can cause productive ecosystems to transition to less productive ones \cite{prod1, prod2, prod3}. Understanding these dynamics is crucial to inform conservation strategies and accurately assess terrestrial ecosystem health. However, predicting state transitions can be difficult—communities of primary producers, which help drive ecosystem dynamics, often exhibit non-deterministic, non-linear, and non-monotonic transition behaviors \cite{markov, non1, non2}.

Biological soil crusts are communities of cyanobacteria, lichen, and bryophytes that appear ubiquitously in terrestrial ecosystems. As dominant primary producers in drylands and tundra, biocrusts account for 7 percent of terrestrial primary productivity and over 50 percent of biological nitrogen fixation \cite{elbert}.

Biocrusts also form assemblages of extracellular secretions and biomass filaments that regulate local hydrological cycling and channel resources from nitrogen-enriched, hydrated biocrusts to depleted plants \cite{assemblage,fl1, fl2, fl3}. These products rehabilitate damaged soil, forming the first stage of colonization in ecosystems recovering from disaster. Indeed, biocrusts are expected to regulate climate impacts on up to 70 percent of global dryland surface \cite{ferrenberg}. Given these ecosystem services and extent of cover, biocrust state transitions are of key consequence.

Ecological succession is the development of community structure over time, with each state characterized by different dominant organisms \cite{eco}. In the initial stage, bare soil is colonized by pioneer cyanobacteria, which bind filamentous sheaths together to stabilize soil microstructure \cite{cyan}. Over time, early stage cyanobacteria are replaced by a more diverse community of late-stage biocrusts, including lichen, cyanolichen, and bryophytes in order of succession. Lichen form from a symbiotic relationship between a photosynthetic partner and fungal components, while mosses perform photosynthesis using stem leaves. Both penetrate soil surfaces through hyphal extension \cite{lichen1, moss2}. Biocrusts perform different ecosystem services depending on their dominant succession stage. For example, moss-dominated biocrusts sequester carbon and dissimilate nitrogen more effectively, while lichen-dominated biocrusts are better suited to metabolize aromatic compounds and exhibit higher nitrogen uptake \cite{moss}. 

Previous research has established that biocrusts are vulnerable to a range of disturbances, including trampling, wildfire, flooding, vehicle traffic, and livestock grazing. Global change may also exert differential effects on biocrust communities based on succession stage, compromising vital ecosystem processes \cite{disturbance}. Several studies have observed the effects of these disturbances on biocrusts. Chronic physical disturbance and natural disaster have been shown to alter community structure for decades, notably accelerating erosion and reducing C and N concentrations in soil \cite{chronic}. More recently, climate manipulation studies have observed reduced biocrust metabolic activity and diminished network connectivity in response to experimental warming \cite{manipulation}. Some work suggests that all types of disturbance trigger biocrust reverse ecological succession, a state transition from late to early stages. For instance, Ferrenberg \textit{et al.} (2014) reported shifts from moss cover to cyanobacteria cover in response to both experimental warming and human trampling \cite{ferrenberg}. Other studies have shown that when subjected to climate manipulation and grazing, early stage cyanobacteria dominate the disturbance gradient, while late-stage cyanobacteria dominate the recovery gradient \cite{gradient1, gradient2}. Prior to reverse succession, disturbed biocrusts may also often undergo dramatic declines in taxonomic diversity \cite{richness}. Though disturbances may vary in the magnitude and rate of reverse succession they trigger, these results suggest that biocrusts exhibit a general, disturbance-agnostic response to suboptimal environmental conditions. The consequences of these induced reverse state transitions can be severe, including deficits in global carbon and nitrogen budgets.

In this study, we develop a time-independent resource-firing game that captures succession dynamics without specifying detailed function forms. Our goals were first, to propose a general model of biocrust succession transitions based on unified theory and second, to examine dynamics in idealized terrestrial ecosystems. For this purpose, we model disturbances as a reduction in available resources, triggering competition between species within the community. The resource-firing game is evaluated on (1) its ability to predict reverse succession in response to a generic disturbance, (2) the episodic and directional character of changes in community structure, and (3) the stability of equilibria.

First, we discuss the theoretical basis of the resource-firing game and show how the model may be implemented computationally. Then, we turn to the characteristics of a biocrust community that has evolved to an equilibrium state, the various dependences of the model, and attacks on the network.

\section{Model}
\subsection{Overview and Theory}
Shifts in succession often exhibit the same profile when subjected to very different environmental disturbances. We compared the effect of climate manipulation treatments, physical disturbance, and disaster events on the relative cover of four biotic groups that dominate biocrusts: early-stage cyanobacteria, late-stage cyanobacteria, lichen, and bryophytes. Relative cover and geographic data from the literature are summarized in Appendix Table \ref{tab:long}. In 47 of the 50 studies in the dataset, biocrusts exhibit a reverse successional response to disturbance or reduced species diversity, irrespective of the type of disturbance. As a result, a key criterion of this model is its ability to predict disturbance-agnostic reverse succession. In each of the 4 studies that did not conform to this trend, moss cover increased in response to warming, physical disturbance, or resource limitation. Notably, this phenomenon appeared in temperate semiarid regions, including Southeast Spain and the Colorado Plateau \cite{rev1, rev2}. After a disturbance, the order of reverse succession in biocrusts follows a predictable pattern: moss, lichen, late-stage cyanobacteria, and early-stage cyanobacteria. However, when biocrusts recover from disturbance, the order of recolonization is often reversed \cite{general1, general2}. We suggest that while different disturbances may vary in the size and frequency of their effects on community structure, succession can arise from the internal dynamics of community-wide resource allocation, rather than the triggering stressor. The resource-firing game proposed in this paper captures these disturbance-agnostic internal dynamics.

Each study in Appendix Table \ref{tab:long} observes biocrust cover both prior to disturbance and after disturbance, where $t_{obs}$ is the elapsed observation period. However, few have captured intermediate succession dynamics. The available literature is sufficient to supply initial conditions and suggest equilibria, but cannot robustly constrain model trajectories at $t < t_{obs}$. Therefore, we preliminarily assume that state transition behaviours proceed monotonically and call for field work to addresses this gap in existing literature. To build our model assumptions from this dataset, we do not specify function forms of relative cover change, instead focusing on macroscopic changes in succession stage. The strategy of biocrust ecological succession is estimated as the attempt to optimally distribute finite resources throughout an interconnected biomass system. This objective is achieved by considering the entire biomass as a nexus of species and attempting to allocating resources based on their fitness \cite{odum}.

\subsection{Order Theory}
Biocrust ecological succession is the development of its community structure over time, with each succession stage characterized by different dominant organisms. In the initial stage, bare soil is colonized by pioneer cyanobacteria. Over time, early stage cyanobacteria are replaced by late-stage biocrusts, including lichen, cyanolichen, and bryophytes in order of succession. We formalize this successional sequence as the partially ordered set $S$. Let $S$ be the chain
\begin{align}
    \centering
    \text{\textit{cyanobacteria}} \longrightarrow \text{\textit{lichen}} \longrightarrow \text{\textit{bryophytes}}
\end{align}

Pioneer early succession populations require fewer resources and thrive in resource-limited conditions, whereas late succession populations outcompete the former in resource-abundant conditions \cite{general1}. Let $\phi$ be a function that maps dominant succession stage to the resources it requires to sustain itself. The notion of resources, while highly abstract, might be captured by applying the species-area relationship, generalized fertility indices, or net primary productivity, the details of which are beyond the scope of this study.

\begin{equation}
\phi(s) = 
    \begin{cases}
        r_{f} & \text{if \textit{s} = fungi} \\
        r_{c} & \text{if \textit{s} = cyanobacteria} \\
        r_{l} & \text{if \textit{s} = lichen} \\
        r_{b} & \text{if\textit{ s} = bryophytes}
    \end{cases}
\end{equation}\label{eqn:phi}

 such that $r_f \leq r_c \leq r_l \leq r_b$. We now turn to the order structure of our game. Explicit $r_f$, $r_c$, $r_l$, and $r_b$ values depend on (1) total resources $R$ available to a community and (2) the number of species of each succession stage $s_f, s_c, s_l, s_b$. The equation for resource allocation is given by
\begin{equation} \label{e:resources}
    \left[\begin{array}{c}
    s_f \\
    s_c \\
    s_l \\
    s_b
\end{array}\right]^T
    \left[\begin{array}{c}
    r_f \\
    r_c \\
    r_l \\
    r_b
\end{array}\right]
= R
\end{equation}

\begin{definition} 
Formally, two partially ordered sets $S$ and $Q$ are \textbf{order-isomorphic }($S \cong Q$) if there exists a map $\phi$ from $S$ onto $Q$ such that $x \leq y$ in $S$ $\iff$ $\phi(x) \leq \phi(y)$ in Q \cite{order}. 
\end{definition}

That is, $\phi$ must faithfully mirror the order structure of $S$. Let $Q = \phi(s)$ be a partially ordered set of allocated resources for $\phi$ defined in \ref{eqn:phi}. Then, $S$ and $Q$ are order-isomorphic, allowing us to directly map the successional composition of a community to its resource allocation. 
\begin{align}
    \centering
    \phi(\text{\textit{cyanobacteria}}) \longrightarrow \phi(\text{\textit{lichen}}) \longrightarrow \phi(\text{\textit{bryophytes}})
\end{align}

\subsection{Representing Biocrust Networks}
\begin{figure}[!htbp]  
\centering 
\begin{tikzpicture}[scale = 0.7]      
\node [label=bare soil,draw,fill=white] (node1) {};
\node [label={[name=l] fungi},draw,fill=brown] (node2) at ([xshift=2cm]node1.east){};  
\node [label={[name=l] early cyan.},draw,fill=yellow] (node3) at ([xshift=2cm]node2.east){};  
\node [label={[name=l] late cyan.},draw,fill=lime] (node4) at ([xshift=2cm]node3.east){};  
\node [label={[name=l] lichen},draw,fill=green] (node5) at ([xshift=2cm]node4.east){};  
\node [label={[name=l] moss},draw,fill=blue] (node6) at ([xshift=2cm]node5.east){};  
\end{tikzpicture}

\begin{tikzpicture}[scale = 0.7]
\graph [typeset=(\tikzgraphnodetext), math nodes, nodes={draw, circle}, n=8, radius=2cm, clockwise]
{1[as=$ $,fill=white,opacity = 1]; 
2[as=$ $,fill=brown,opacity = 1]; 
3[as=$ $,fill=brown,opacity = 1]; 
4[as=$ $,fill=lime,opacity = 1]; 
5[as=$ $,fill=blue,opacity = 1]; 
6[as=$ $,fill=lime,opacity = 1]; 
7[as=$ $,fill=green,opacity = 1]; 
8[as=$ $,fill=white,opacity = 1]; 
subgraph K_n };
\end{tikzpicture}
\begin{tikzpicture}[scale = 0.7]
\graph [typeset=(\tikzgraphnodetext), math nodes, nodes={draw, circle}, n=12, radius=2cm, clockwise]
{1[as=$ $,fill=white,opacity = 1]; 
2[as=$ $,fill=lime,opacity = 1]; 
3[as=$ $,fill=yellow,opacity = 1]; 
4[as=$ $,fill=lime,opacity = 1]; 
5[as=$ $,fill=blue,opacity = 1]; 
6[as=$ $,fill=lime,opacity = 1]; 
7[as=$ $,fill=green,opacity = 1]; 
8[as=$ $,fill=brown,opacity = 1]; 
9[as=$ $,fill=white,opacity = 1]; 
10[as=$ $,fill=yellow,opacity = 1]; 
11[as=$ $,fill=lime,opacity = 1]; 
12[as=$ $,fill=green,opacity = 1]; 
subgraph K_n };
\end{tikzpicture}
\caption{
\textbf{Biocrust networks of varying scale.} Both graphs are complete, with 8 and 12 species respectively (Color Required)
} \label{fig:M}
\end{figure}
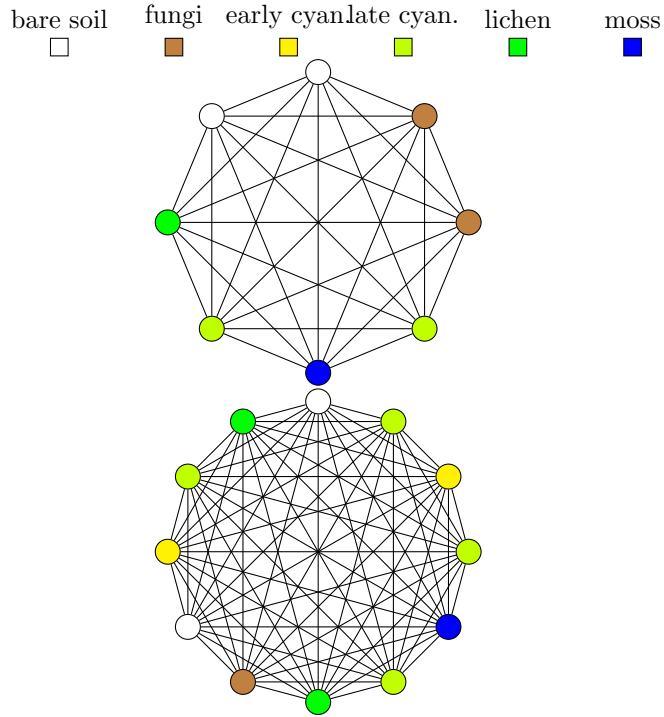

A biocrust community is modeled as an interconnected network that distributes resources to several populations of species competing to maximize ecological fitness. The strategy of biocrust ecological succession is estimated as the attempt to optimally distribute finite resources throughout an interconnected biomass \cite{odum}.  
\begin{definition}
   A \textbf{complete graph} is a simple undirected graph $G$ in which every pair of nodes is connected by a unique edge.
\end{definition}
We represent a generic biocrust network as a complete graph. A complete graph with $n$ nodes has $\frac{n(n-1)}{2}$ edges. Each node represents a species in the community, and an edge represents competition for shared resources. In other words, each species in the biocrust community is assumed to be in competition with every other species. Each node has two fields. The first field is its designated succession stage, $s$: fungus, early cyanobacteria, late cyanobacteria, lichen, or bryophyte. The second field is the concomitant number of resources, $\phi(s)$. The succession stage with a plurality of designated nodes is said to dominate the community. Figure \ref{fig:M} shows two complete graphs, representing a fungi- and late cyanobacteria-dominated community with 8 species and a late cyanobacteria-dominated community with 12 species. The following section describes the rules of the resource-firing game on such graphs.

\subsection{Resource-Firing Game}\label{section:model}
The chip-firing game models the loss of energy in networks of individuals whose states depend on the state of their neighbors. In this paper, we define the \textbf{resource-firing game}, a portmanteau of chip-firing game for resource allocation. The game is executed on a finite graph $G = (V,E)$ with $|V| = n$ \textbf{nodes}, in addition to a \textbf{sink node}, which becomes active when every species in the community is depleted of resources \cite{klivans}.

Given any biocrust network, we can play the resource allocation game as follows. First, we add a sink node to $G$. For $G$ to remain complete, the sink node must have edges to every species in the network. Thus, every community $G$ has $n$ nodes, 1 sink node and $n-1$ species, retaining the property of completeness. Next, a non-negative number of resources are distributed to each node, with the exception of the sink node. Recall that the distribution of resources to each species as a function of its succession stage is given by Equation \ref{e:resources}. We write the resource configuration as follows: 

\begin{equation}
    \mathbf{r} = \{r_1,r_2,...,r_k\} \in \mathds{N}_0
\end{equation}

\noindent where $r_k$ is the number of resources on node $n_k$. The basic step of this game is firing initiated by a generic disturbance, the dynamics of which are detailed in Algorithm \ref{alg:cap}. Firing represents a transfer of resources from one stressed species to its non-stressed neighbors. Accordingly, the node with highest $r_n$, equivalently lowest fitness, fires. 
%\begin{definition}
%The \textbf{degree} of node $n_k$ ($deg(n_k)$) is the number of nodes $n_k$ has edges to, or equivalently neighbors.
%\end{definition}

A node $n_k$ can be fired if $r_k \geq deg(n_k)$, where $deg(n_k)$ is equivalent to the number of neighbors $n_k$ has. For example, if $n_1$ has edges to $n_2, n_4,$ and $n_5$, those nodes are considered its neighbors and $deg(n_1) = 3$. The set of neighbor pairs is denoted $E(G)$. When node $n_k$ fires, it sends 1 chip to each of its neighbors, losing $deg(n_k)$ resources. Biologically, this indicates that the species occupying $n_k$ no longer has sufficient resources to sustain itself. The released resources now become available to competing species. If $n_1$ is selected to fire and initially has $5$ resources, it sends $1$ resource to each its $3$ neighbors and retains $2$ resources. As a result, the succession stage of the species occupying $n_1$ regresses to one that can be supported by 2 resources, a process which we will call reverse succession. Dually, $n_2, n_4,$ and $n_5$, $n_1$'s neighbors, gain resources and can support species with advanced succession stages, a process which we will call forward succession. The mechanisms of forward and reverse succession are beyond the scope of the present manuscript, and we refer readers to the work of others. Formally, the piece-wise function $r'(i)$ denotes the successor configuration of of any node $i$ in the graph after a given node $j$ fires:

\begin{equation}
\mathbf{r'(i)} =
    \begin{cases}
        r_j - deg(n_j) & \text{if } i = j\\
        r_i + 1 & \text{if }(n_i,n_j) \in E(G)\\
        r_i & \text{if } (n_i,n_j) \notin E(G)
    \end{cases}
\end{equation} 

Redistribution of resources after firing may enable new non-sink nodes to fire. The game ends when the community arrives at a \textbf{stable configuration} in which no nodes are eligible to fire. At this point, the sink vertex releases its chips to each of its neighbors and the game restarts. The length of the game is defined as the number of firings until a stable configuration is reached. 

Because chip firing exhibits terminating behavior, the stable configuration is always unique. Chip-firing also displays local confluence, meaning nodes can be fired in any order without affecting the length or final configuration of the game \cite{confluence}.

\begin{algorithm}[tb]
\caption{Resource-Firing Algorithm}\label{alg:cap}
\begin{algorithmic}
\Require $s$ is \textbf{sink node}, $V \neq \emptyset$ and contains all nodes except $s$
\Ensure $t_{limit}$ iterations of resource-firing on a graph with $n$ nodes and $r$ resources

\While{$t < t_{limit}$}
\State $n_{max} = \{n_{i} \in V$ $|$ $\forall n_{j} \in V \quad$ $r_{j} \leq r_{i}\}$ \\
\If{$r_{max} \geq deg(n_{max})$} 
    \State $r_{max} = r_{max} - deg(n_{max})$ 
    \State $ \forall (n_i, n_j) \in E(G) \quad$ $r_{j} = r_{j} + 1$
\Else  
    \State $r_{sink} = 0$     
    \State $\forall n_{j} \in V \quad$  $r_j = r_j + 1$
\EndIf 
\EndWhile
\end{algorithmic}
\end{algorithm}

\section{Dynamics in Terrestrial Ecosystems}
\subsection{Example: $n = 5$}

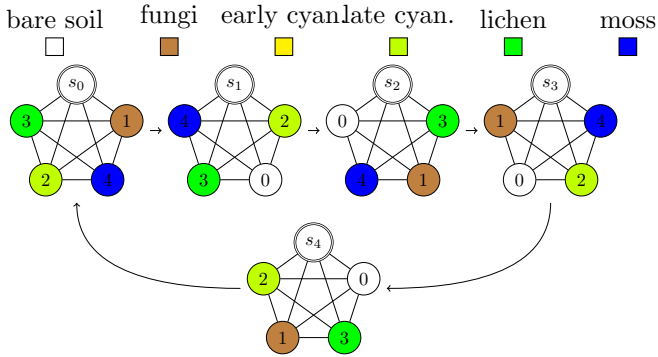
\begin{figure}[H]
\centering
\begin{tikzpicture}[scale = 0.7]       
\node [label=bare soil,draw,fill=white] (node1) {};
\node [label={[name=l] fungi},draw,fill=brown] (node2) at ([xshift=2cm]node1.east){};  
\node [label={[name=l] early cyan.},draw,fill=yellow] (node3) at ([xshift=2cm]node2.east){};  
\node [label={[name=l] late cyan.},draw,fill=lime] (node4) at ([xshift=2cm]node3.east){};  
\node [label={[name=l] lichen},draw,fill=green] (node5) at ([xshift=2cm]node4.east){};  
\node [label={[name=l] moss},draw,fill=blue] (node6) at ([xshift=2cm]node5.east){};  
\end{tikzpicture}
\scalebox{0.7}{
\begin{tikzpicture}[node distance=4cm]
\node (a) at (0,0) {\begin{tikzpicture}[scale = 0.6]
\graph [typeset=(\tikzgraphnodetext), math nodes, nodes={draw, circle}, n=5, radius=1cm, clockwise]
{1[as=$s_0$,fill=white,opacity = 1,accepting]; 2[as=$1$,fill=brown,opacity = 1]; 3[as=$4$,fill=blue,opacity = 1]; 4[as=$2$, fill=lime,opacity = 1]; 5[as=$3$,fill=green,opacity = 1]; subgraph K_n };
\end{tikzpicture}};
\node (b) at (3,0) {\begin{tikzpicture}[scale = 0.6]
\graph [math nodes, nodes={draw, circle}, n=5, radius=1cm, clockwise]
{1[as=$s_1$,fill=white,opacity = 1,accepting]; 2[as=$2$,fill=lime,opacity = 1]; 3[as=$0$,fill=white,opacity = 1]; 4[as=$3$,fill=green,opacity = 1]; 5[as=$4$,fill=blue,opacity = 1]; subgraph K_n };
\end{tikzpicture}};
\node (c) at (6,0) {\begin{tikzpicture}[scale = 0.6]
\graph [math nodes, nodes={draw, circle}, n=5, radius=1cm, clockwise]
{1[as=$s_2$,fill=white,opacity = 1,accepting]; 2[as=$3$,fill=green,opacity = 1]; 3[as=$1$,fill=brown,opacity = 1]; 4[as=$4$,fill=blue,opacity = 1]; 5[as=$0$,fill=white,opacity = 1]; subgraph K_n };
\end{tikzpicture}};
\node (d) at (9,0) {\begin{tikzpicture}[scale = 0.6]
\graph [math nodes, nodes={draw, circle}, n=5, radius=1cm, clockwise]
{1[as=$s_3$,fill=white,opacity = 1,accepting]; 2[as=$4$,fill=blue,opacity = 1]; 3[as=$2$,fill=lime,opacity = 1]; 4[as=$0$,fill=white,opacity = 1]; 5[as=$1$,fill=brown,opacity = 1]; subgraph K_n };
\end{tikzpicture}};
\node (e) at (4.5,-3.0) {\begin{tikzpicture}[scale = 0.6]
\graph [math nodes, nodes={draw, circle}, n=5, radius=1cm, clockwise]
{1[as=$s_4$,fill=white,opacity = 1,accepting]; 2[as=$0$,fill=white,opacity = 1]; 3[as=$3$,fill=green,opacity = 1]; 4[as=$1$,fill=brown,opacity = 1]; 5[as=$2$,fill=lime,opacity = 1]; subgraph K_n };
\end{tikzpicture}};

\draw[->] (a) -- node[above] {} (b);
\draw[->] (b) -- node[above] {} (c);
\draw[->] (c) -- node[above] {} (d);
\draw[ ->] (d.south) to [out=270,in=0] (e.east);
\draw[ ->] (e.west) to [out=180,in=270] (a.south);
\end{tikzpicture}
}
\caption{ \textbf{Sequence of chip firings} on a biocrust community with $n=5$ node and 10 resources. The sink nodes in each configuration are denoted $s_0,s_1, s_2, s_3, s_4$, respectively. Graph 5 represents the stable configuration} \label{fig:E}
\end{figure}

Figure \ref{fig:E} shows a simple firing sequence on a biocrust community with $n = 5$ nodes (4 species) and $r = 10$ resources. As the sink node accumulates resources and the number of resources available to the community decreases, it undergoes reverse succession. Conversely, the release of resources from $s$ triggers forward succession, yielding the initial configuration.

The resulting stable configuration appears on the bottom. Mathematically, the stable configuration refers to a configuration of resources in which each node has at most as many chips as its degree. Furthermore, there is no subset of nodes that can fire. The resource-firing game will always proceed to a stable configuration and remain there forever, irrespective of how nodes fire. In other words, the stable configuration represents an equilibrium or fixed point. 

Now we turn to the stability of the configurations in our $n=5$ example. Let $c_0$ represent the fixed point of $f'(r'(i)) = (r'(1),r'(2),r'(3),r'(4))$, where i is the node index. Given that $f^{(n)} \rightarrow c_0$, we conclude that $c_0$ is a stable fixed point. For the given initial configuration, $f$ has a four-cycle,

\begin{equation}
f^{(4)}(c_0) = c_0, f'(c_0) = f''(c_0) = f^{(3)}(c_0) \neq c_0.
\end{equation}

\noindent We might also attempt to determine the stability of $c_0$ by using the Tutte polynomial.

\begin{definition}
    For an undirected graph $G=(V,E)$, the \textbf{Tutte polynomial} is $T_{G}(x,y)=\sum \nolimits _{{A\subseteq E}}(x-1)^{{k(A)-k(E)}}(y-1)^{{k(A)+|A|-|V|}}$, where $k(A)$ denotes the number of connected components of the graph $(V,A)$ \cite{tutte}.
\end{definition}

\begin{definition}
    A \textbf{minimum spanning tree} is a tree that includes every node of a given graph and the minimal set of edges needed to connect each node without cycles \cite{mst}.
\end{definition}

\begin{figure}[!htbp]  
\centering 
\begin{tikzpicture}
\begin{axis}[
  grid=major,              % draw major gridlines
  major grid style=dotted, % dotted grid lines
  colormap/jet,            % colormap from MATLAB
  samples=30,              % 30 samples in each direction
  view={140}{30},          % configure plot view
  domain=-3:3,             % x varies from -3 to 3
  y domain=-3:3,           % y varies from -3 to 3
  zmin=-10, zmax=30,       % z-axis limits
  xlabel={$x$},            % x-axis label
  xtick={-3,-2,...,3},     % integer-spaced tick marks on the x-axis
  ylabel={$y$},            % y-axis label
  title={$x^3 + y^3 + 3x^2 + 4xy + 3y^2 + 2x + 2y$},     % plot title
]
  \addplot3[mesh] {x^3 + y^3 + 3*x^2 + 4*x*y + 3*y^2 + 2*x + 2*y}; % make the mesh plot
\end{axis}
\end{tikzpicture}
\caption{\textbf{Tutte Polynomial} of the $n=5$ game shown in Figure \ref{fig:E}} \label{fig:tutte}
\end{figure}
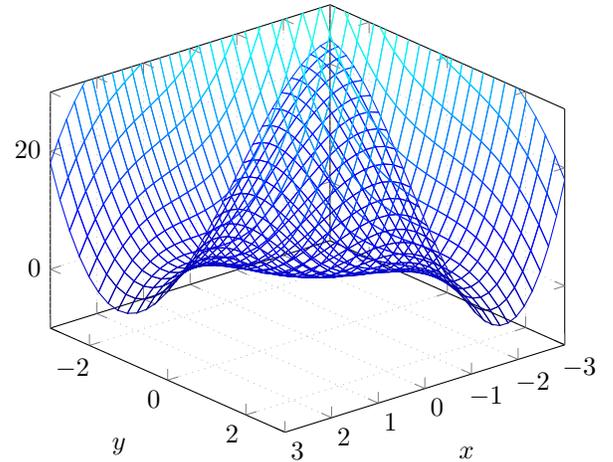

\noindent 
Evaluating the Tutte polynomial of a graph $T_G$ at $T_G(1,1)$ yields the number of spanning trees, which provides information about stable configurations. \cite{tutte1}. The Tutte polynomial of $G_5$ in Figure \ref{fig:E} is given in Figure \ref{fig:tutte}. Note that the coefficients are non-negative.

\subsection{Species Richness Dependence}

In the following two sections, we consider additive factors to the basic model introduced in Section \ref{section:model}. We begin with a first approximation of the species richness of global terrestrial ecosystems compiled from available literature data in Appendix Table \ref{tab:long}.

\begin{table}[!htbp]
%\tbl{Idealized Species Richness by Ecosystem}
\centering
{\begin{tabular}{lcccccc} \toprule
% Ecosystem & Species \\ \toprule
 Desert & $0<n<30$ \\
 Steppe & $30<n<50$\\
 Temperate Forest & $50<n<100$ \\ \bottomrule
\end{tabular}}
\label{table1}
\end{table}

 First, we examine how the dependence of the resource-firing on the number of species ($n$) affects game stability. A common approach derived from topological graph theory to measure chip-firing stability is to compute the genus of the graph.

\begin{definition}
    The \textbf{genus} of a graph is the minimum integer $k$ such that the graph can be topologically embedded onto a sphere with $k$ handles without crossing itself \cite{genus1}.
\end{definition}

\begin{figure}[!htbp]  
\centering 
\begin{tikzpicture}[scale = 0.8]
\begin{axis}[
  only marks,                    % no lines
  xmin=0, xmax=100,              % x-axis limits
  ymin=0, ymax=5005,              % y-axis limits
  xlabel={Species},      % x-axis label
  ylabel={Genus},            % y-axis label
  %title={Genus by Ecosystem}, % plot title
  legend pos=north west,         % legend position on plot
  legend cell align=left,        % text alignment within legend
  domain=20:180,                 % domain for plotted functions (not needed for scatter data)
  samples=200,                   % plot 200 samples
]
  \fill[blue!20] (0,0) rectangle (32,5000);     % fill a blue rectangle behind the blue data
  \fill[red!20] (32,0) rectangle (52,5000);     % fill a red rectangle behind the red data
  \fill[green!20] (52,0) rectangle (102,5005);  % fill a green rectangle behind the green data
  \addplot[mark=o,blue] coordinates {
		(1,0)
            (5,6)
            (10,36)
            (15,91)
            (20,171)
            (25,276)
            (30,406)  
	};
  \addlegendentry{Desert}; 
  \addplot[mark=+,red]  coordinates {
            (35,561)
            (40,741)
            (45,946)
            (50,1176)
  };
  \addlegendentry{Steppe};
  \addplot[mark=triangle,teal] coordinates {
            (55,1431)
            (60,1711)
            (65,2016)
            (70,2346)
            (75,2701)
            (80,3081)
            (85,3486)
            (90,3916)
            (95,4371)
            (100,4851)
  };
  \addlegendentry{Temperate Forest};
\end{axis}
\end{tikzpicture}
\caption{Genus by Ecosystem (Color Required)} \label{fig:G}
\end{figure}
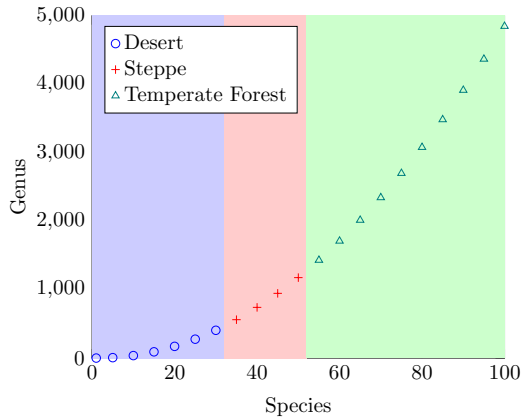

\noindent If the graph embedding of $G$ is topologically simple, $G$ has a low genus and the game may likely requires fewer resources to be available at the outset. Dually, if $G$ has a high genus, the game may require a larger number of resources \cite{genus1, genus2}. In Figure \ref{fig:G}, we compute the genus of resource-firing games for $0 < n < 100$ for idealized desert, steppe, and temperate forest biocrust communities. One consequence of this analysis is that genus increases quadratically with respect to $n$. Ergo, the resource stability threshold of idealized forest communities may be significantly higher than those of desert and semi-desert communities.

\begin{figure}[!htbp]  
\centering 
\begin{tikzpicture}[scale = 0.8]
\begin{axis}[
  only marks,                    % no lines
  xmin=0, xmax=5,              % x-axis limits
  ymin=0, ymax=10,              % y-axis limits
  xlabel={$\log(N)$},      % x-axis label
  ylabel={$\log(P(N))$},            % y-axis label
  %title={Avalanche Distribution by Ecosystem}, % plot title
  legend pos=north west,         % legend position on plot
  legend cell align=left,        % text alignment within legend
  samples=200,                   % plot 200 samples
]
  \addplot[mark=o,blue] coordinates {
            (0,9.2)
            (0.7,5.6)
            (1.2,4.9)
            (1.4,4.4)
            (1.6,4)
            (1.8,3.6)
            (1.95,3.42)
            (2.066,2.9)
            (2.22,3.5)
            (2.3,3.4)
            (2.4,3.6)
            (2.5,4)
            (2.6,4.3)
            (2.7,5.6)         
  }; % ...
  \addlegendentry{Desert ($n=15$)};
  \addplot[mark=+,red] coordinates {
            (0.7,4.4)
            (1.1,3.7)
            (1.4,3.2)
            (1.6,2.9)
            (1.8,2.7)
            (1.95,2)
            (2.1,2.3)
            (2.2,1.5)
            (2.3,1.5)
            (2.566,0)
            (3.033,0)
            (3.333,1.6)
            (3.366,0)
            (3.45,2.1)
            (3.5,1.95)
            (3.55,2.15)
            (3.57,2.8)
            (3.62,2.8)
            (3.6,2.2)
            (3.65,2.45)
            (3.67,3.2)
            (3.7,4)
            (3.75,5)
  };
  \addlegendentry{Steppe ($n=40$)};
  \addplot[mark=triangle,teal] coordinates {
            (0,8)
            (0.7,4)
            (1.1,2.8)
            (1.4,2.5)
            (1.6,1.8)
            (1.75, 0)
            (1.92,0)
            (2.08,0)
            (4.18,0)
            (4.2,0)
            (4.22,0)
            (4.25,0.6)
            (4.31,2.4)
            (4.33,2.6)
            (4.35,3.2)
            (4.36,3.53)
            (4.38,4.4)
  }; % add the first plot
  \addlegendentry{Temperate Forest ($n=75$)}; % add the first plot's legend entry
\end{axis}
\end{tikzpicture}
\caption{Avalanche Distribution by Ecosystem (Color Required)} \label{fig:aval}
\end{figure}
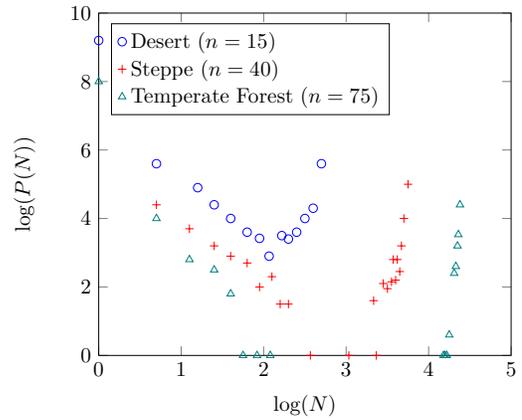

Next, we consider how species richness impacts the avalanche distribution of the resource-firing game. When a node fires, a stressed species releases resources to non-stressed neighbors. This reallocation of resources may enable other species to release resources, resulting a percolation of firings across the graph. An avalanche is precisely this trickling sequence of firings. The length of the avalanche is the length of the sequence, that is, the number of firings that occur until firing must be induced for the game to proceed. The avalanche distribution is the probability distribution of the lengths of avalanches that occur, given some initial configuration of chips. For idealized desert, steppe, and temperate forest communities with $n=15$, $n=40$, and $n=75$ species respectively, the resource-firing game was initiated 10,000 times and the avalanche distribution in Figure \ref{fig:aval} generated.

The distribution suggests that reverse succession is both episodic and directional, proceeding towards reverse succession in bursts of variable length. In all ecosystems, avalanches are frequently short or long, but not often medium-sized. As $n$ increases, two effects are observed: the probability of medium-sized avalanches tends towards zero and the average length of avalanches increases. 

\subsection{Attacks on Biocrust Communities}
The resource-firing game assumes a uniformly competitive interaction between species. In reality, species compete most aggressively with species in their proximity for local pools of resources. For a complete graph $G$ representing a biocrust community, we define an accompanying path graph—a Cayley tree—that approximates the spatial structure of the community. By Cayley's formula, for $G$ with $n$ nodes, $n^{n-2}$ unique spanning trees exist \cite{ct}. To build our Biocrust Cayley Tree (BCT), we replicate all $n$ nodes on $G$, then place edges on them such that a desired spanning tree is generated. The structure of the BCT is as follows. There is a singular node $v$ located at the root at the tree with edges to $k$ neighbors. These neighbors comprise the first shell of the tree. Each node in the first shell has $k-1$ neighbors, forming the second shell. Each node in every subsequent shell also has $k-1$ neighbors \cite{amikam}.

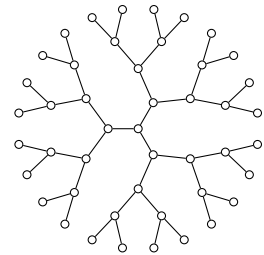
\begin{figure}[!htbp]
    \centering
\begin{tikzpicture}[scale = 0.4]
  \draw (0,0) node[disc=c-0-1];
  \xdef\radius{0cm}
  \xdef\level{0}
  \xdef\nbnodes{1}
  \xdef\degree{(3+1)} % special degree just for the root node
  \foreach \ndegree/\form in {3/disc,3/disc,3/disc,3/disc}{
    \pgfmathsetmacro\nlevel{int(\level+1)}
    \pgfmathsetmacro\nnbnodes{int(\nbnodes*(\degree-1))}
    \pgfmathsetmacro\nradius{\radius+1cm}
    \draw[white] (c-0-1) circle(\nradius pt);
    \foreach \div in {1,...,\nnbnodes} {
      \pgfmathtruncatemacro\src{((\div+\degree-2)/(\degree-1))}
      \path (c-0-1) ++({\div*(360/\nnbnodes)-180/\nnbnodes}:\nradius pt) node[\form=c-\nlevel-\div];
      \draw (c-\level-\src) -- (c-\nlevel-\div);
    }
    \xdef\radius{\nradius}
    \xdef\level{\nlevel}
    \xdef\nbnodes{\nnbnodes}
    \xdef\degree{\ndegree}
  }
\end{tikzpicture}
    \caption{Cayley Tree with $n=45$ and $k = 3$}
    \label{fig:cayley}
\end{figure}

A random attack on the BCT compromises community structure by arbitrarily removing one node at a time. Attacks represent critical disturbances, such as climate change, trampling, or altered precipitation. Three types of nodes are defined: alive, incapacitated, and dead. If a node is attacked, the species it represents dies, and as the attack percolates through the community, its descendants are critically incapacitated. The species represented by all other nodes are considered alive. Let $p_A = \frac{n_A}{n}$ be the fraction of alive species in the community, where $n$ is the total number of nodes in the BCT. We define $A$ as the event that the species selected for attack is alive and $D$ as the event that it is already dead. Given layers $L$ and the number of neighbors $k$, Equation \ref{eqn:PA} approximates the conditional probability of $A$ given $\bar{D}$ \cite{amikam}. Note that a very large $L$ is assumed.

\begin{equation}\label{eqn:PA}
    P(A|\bar{D}) = \frac{(k-2)(1-p_A)^{L+1}}{(k-2) - (k-1)p_A}
\end{equation}

Based on their degree of severity, critical disturbances can lead to species death events of various sizes. We denote such events $E$. We estimate the probability of $E$ with a death toll of at least $k$ as 
\begin{equation}
    P(E) = \frac{L(k-1)}{k-2} \cdot P(L_D)
\end{equation}
where $P(L_D)$ is the probability that the disturbance is severe enough to attack the BCT, that is, cause species death \cite{amikam}.

\section{Conclusions}
We have presented the resource-firing game as a model of resource allocation in complex networks, with a particular focus on its application to biocrust communities. First, the model was motivated by experimental results. Then, we discussed its theoretical basis and the characteristics of a biocrust community that has evolved to an equilibrium state. 

We note here that the initial formulation of the resource-firing game hinges on some simplifying assumptions. For example, it assumes species are equally likely to compete with one another. In real-world ecosystems, species interaction also depends on differences in behavior or spatial proximity. Additionally, the model assumes that interactions are either present or absent, though they may exist on a spectrum of strength or frequency. However, the simple rules of our formulation adequately predict reverse succession in response to a generic disturbance, the episodic and directional character of changes in community structure, and the stability of equilibria.

One key result of our analysis is that changes in species richness have nonlinear effects on network structure and the avalanche distribution of the resource-firing game. In particular, the resource stability threshold for idealized forest communities is significantly higher than that of desert and semi-desert communities, and high species richness increases the probability of very short and long avalanches, but not those of intermediate length. These results suggests that the community’s response to disturbance is both directional and episodic, proceeding towards reverse succession in bursts of variable length. 

We also introduce the BCT as an approximation of the spatial structure of a biocrust community and derive probabilities of critical disturbances that can compromise community structure.

\begin{table*}[t]
\centering
\begin{tabular}{l|l|l|l}
\toprule%
\small{Reference} & \small{Ecosystem Type} & \small{Disturbance} & \small{Data Reported} \\
\midrule
\footnotesize{Campbell (1979) \cite{campbell}}  & \footnotesize{Semi-desert} & \footnotesize{Sand burial} & \footnotesize{Other parameters} \\
\footnotesize{Johansen \textit{et al.} (1984) \cite {joh}} & \footnotesize{Desert} & \footnotesize{Fire} & \footnotesize{Relative cover} \\
\footnotesize{Johansen and St. Clair (1986) \cite{stc}} & \footnotesize{Desert} & \footnotesize{Grazing} & \footnotesize{Relative cover} \\
\footnotesize{Harper and Marble (1988) \cite{harp}} & \footnotesize{--} & \footnotesize{Vascular Plants} & \footnotesize{General trends} \\
\footnotesize{Johansen \textit{et al.} (1993) \cite{joh1}} & \footnotesize{Steppe} & \footnotesize{Fire} & \footnotesize{Species richness} \\
\footnotesize{Kaltenecker \textit{et al.} (1997) \cite{kal}} & \footnotesize{Steppe} & \footnotesize{Fire} & \footnotesize{Species richness} \\
\footnotesize{Eldridge and Koen (1998) \cite{eld}} & \footnotesize{Desert} & \footnotesize{Trampling, Fire} & \footnotesize{Relative cover} \\

\footnotesize{Belnap and Phillips (2001) \cite{bel}} & \footnotesize{Grassland} & \footnotesize{Grass invasion} & \footnotesize{Species richness} \\

\footnotesize{Belnap and Eldridge (2001) \cite{general1}} & \footnotesize{--} & \footnotesize{--} & \footnotesize{General trends} \\

\footnotesize{Belnap and Lange (2001) \cite{bel2}} & \footnotesize{--} & \footnotesize{ Frequent fires} & \footnotesize{General trends} \\

\footnotesize{Belnap \textit{et al.} (2004) \cite{bel3}} & \footnotesize{Desert} & \footnotesize{Rain} & \footnotesize{Other parameters} \\

\footnotesize{Myers and Davis (2003) \cite{myers}} & \footnotesize{Temperate forest} & \footnotesize{ Fire} & \footnotesize{Species richness} \\

\footnotesize{Hilty \textit{et al.} (2004) \cite{hilty}} & \footnotesize{Steppe} & \footnotesize{Fire} & \footnotesize{Species richness} \\

\footnotesize{Bowker \textit{et al.} (2005) \cite{bowker}} & \footnotesize{Semi-desert} & \footnotesize{Micronutrients} & \footnotesize{Relative cover} \\

\footnotesize{Thompson \textit{et al.} (2005) \cite{thompson}} & \footnotesize{Desert} & \footnotesize{Soil fertility} & \footnotesize{Relative cover} \\

\footnotesize{Belnap \textit{et al.} (2006) \cite{bel4}} & \footnotesize{Semi-desert} & \footnotesize{Grass invasion} & \footnotesize{Relative cover} \\

\footnotesize{Callaway (2007) \cite{callaway}} & \footnotesize{Desert} & \footnotesize{Micronutrients} & \footnotesize{Species richness} \\

\footnotesize{Williams \textit{et al.} (2008) \cite{williams}} & \footnotesize{Desert} & \footnotesize{Grazing, drought} & \footnotesize{Relative cover} \\

\footnotesize{Budel \textit{et al.} (2008) \cite{budel}} & \footnotesize{Desert} & \footnotesize{Trampling, Rain} & \footnotesize{Other parameters} \\

\footnotesize{Lazaro \textit{et al.} (2008) \cite{lazaro}} & \footnotesize{Semi-desert} & \footnotesize{Micronutrients} & \footnotesize{Other parameters} \\

\footnotesize{Kidron \textit{et al.} (2008) \cite{kidron}} & \footnotesize{Desert} & \footnotesize{Grazing} & \footnotesize{Other parameters} \\

\footnotesize{Wang \textit{et al.} (2009) \cite{wang}} & \footnotesize{Desert} & \footnotesize{Sand burial} & \footnotesize{Relative cover} \\

\footnotesize{Li \textit{et al.} (2010) \cite{li1}} & \footnotesize{Desert} & \footnotesize{Geomorphology} & \footnotesize{Other parameters} \\

\footnotesize{Langhans \textit{et al.} (2010) \cite{langhans}} & \footnotesize{Temperate forest} & \footnotesize{Fine-disturbance} & \footnotesize{General trends} \\

\footnotesize{Kuske \textit{et al.} (2011) \cite{chronic}} & \footnotesize{Semi-desert} & \footnotesize{Trampling} & \footnotesize{Species richness} \\

\footnotesize{Li \textit{et al.} (2011) \cite{li2}} & \footnotesize{Desert} & \footnotesize{Ant nests} & \footnotesize{Other parameters} \\

\footnotesize{Escolar \textit{et al.} (2012) \cite{rev2}} & \footnotesize{Desert} & \footnotesize{Warming, Rain} & \footnotesize{Relative cover} \\

\footnotesize{Reed \textit{et al.} (2012) \cite{reed}} & \footnotesize{Semi-desert} & \footnotesize{Rain} & \footnotesize{Relative cover} \\

\footnotesize{Johnson \textit{et al.} (2012) \cite{johnson}} & \footnotesize{Semi-desert} & \footnotesize{Warming, Rain} & \footnotesize{Other parameters} \\

\footnotesize{Zelikova \textit{et al.} (2012) \cite{zelikova}} & \footnotesize{Semi-desert} & \footnotesize{Warming, Rain} & \footnotesize{Relative cover} \\

\footnotesize{Yeager \textit{et al.} (2012) \cite{yeager}} & \footnotesize{Semi-desert} & \footnotesize{Warming} & \footnotesize{Other parameters} \\

\footnotesize{Maestre \textit{et al.} (2013) \cite{maestre}} & \footnotesize{Semi-desert} & \footnotesize{Warming} & \footnotesize{Relative cover} \\

\footnotesize{Concostrina-Zubiri \textit{et al.} (2013) \cite{cz13}} & \footnotesize{Desert} & \footnotesize{Micronutrients} & \footnotesize{Species richness} \\

\footnotesize{Concostrina-Zubiri \textit{et al.} (2014) \cite{cz14}} & \footnotesize{Desert} & \footnotesize{Grazing} & \footnotesize{Species richness} \\

\footnotesize{Zhang \textit{et al.} (2014) \cite{zhang}} & \footnotesize{Desert} & \footnotesize{Sand burial} & \footnotesize{Other parameters} \\

\footnotesize{Ferrenberg \textit{et al.} (2015) \cite{ferrenberg}} & \footnotesize{Semi-desert} & \footnotesize{Trampling} & \footnotesize{Relative cover} \\

\footnotesize{Maestre \textit{et al.} (2015) \cite{maestre1}} & \footnotesize{Desert} & \footnotesize{Warming} & \footnotesize{Relative cover} \\

\footnotesize{Steven \textit{et al.} (2015)\cite{steven}} & \footnotesize{Semi-desert} & \footnotesize{Trampling} & \footnotesize{Species richness} \\

\footnotesize{Budel \textit{et al.} (2016) \cite{budel1}} & \footnotesize{--} & \footnotesize{--} & \footnotesize{General trends} \\

\footnotesize{Belnap \textit{et al.} (2016) \cite{general1}} & \footnotesize{--} & \footnotesize{--} & \footnotesize{General trends} \\

\footnotesize{Rutherford \textit{et al.} (2017) \cite{rutherford}} & \footnotesize{Semi-desert} & \footnotesize{Warming} & \footnotesize{Relative cover} \\

\footnotesize{Fernandes \textit{et al.} (2017) \cite{fernandes}} & \footnotesize{Desert} & \footnotesize{Rain} & \footnotesize{Species richness} \\

\footnotesize{Li \textit{et al.} (2017) \cite{li3}} & \footnotesize{Desert} & \footnotesize{Rain, Topsoil} & \footnotesize{Species richness} \\

\footnotesize{Zhang \textit{et al.} (2017) \cite{zhang2}} & \footnotesize{Desert} & \footnotesize{Sand burial} & \footnotesize{Species richness} \\

\footnotesize{Eldridge \textit{et al.} (2018) \cite{eldridge2}} & \footnotesize{Desert} & \footnotesize{Warming} & \footnotesize{Species richness} \\

\footnotesize{Ladron de Guevara \textit{et al.} (2018) \cite{ldg}} & \footnotesize{Desert} & \footnotesize{Warming} & \footnotesize{Relative cover} \\

\footnotesize{Li \textit{et al.} (2018) \cite{li4}} & \footnotesize{Desert} & \footnotesize{Warming} & \footnotesize{Relative cover} \\

\footnotesize{Aanderud \textit{et al.} (2019) \cite{aanderud}} & \footnotesize{Desert} & \footnotesize{Fire} & \footnotesize{Species richness} \\

\footnotesize{Berdugo \textit{et al.} (2021) \cite{berdugo}} & \footnotesize{Temperate forest} & \footnotesize{Warming} & \footnotesize{Other parameters} \\

\footnotesize{Li \textit{et al.} (2021) \cite{li5}} & \footnotesize{Desert} & \footnotesize{Warming} & \footnotesize{Relative cover} \\

\footnotesize{Lan \textit{et al.} (2021) \cite{lan}} & \footnotesize{Desert} & \footnotesize{Sand burial} & \footnotesize{Other parameters} \\

\footnotesize{Finger-Higgens \textit{et al.} (2022) \cite{fh}} & \footnotesize{Semi-desert} & \footnotesize{Grass invasion} & \footnotesize{Relative cover} 
\end{tabular}

\caption{Ecological succession of biocrust covers from available literature data (Appendix)}\label{tab:long} 
\end{table*}

\section{Declarations}
\subsection{Acknowledgements}
The author thanks Erwan Monier, who provided feedback on technical aspects of the model.

\subsection{Funding}
The author declares that no funds, grants, or other support were received during the preparation of this manuscript.

\subsection{Competing Interests}
The author declares no competing interests.

\clearpage

\end{document}